\documentclass[runningheads]{llncs}
\usepackage{graphicx}
\usepackage{dcolumn}
\usepackage{endnotes}
\usepackage{siunitx}
\usepackage{marvosym}
\usepackage{cite}

\usepackage{algorithm}  
\usepackage{algpseudocode}  
\usepackage{amsmath}  
\usepackage{hyperref}\hypersetup{
  colorlinks=true,
  linkcolor=blue,
  filecolor=blue,      
  urlcolor=blue,
  citecolor=blue,
}

\begin{document}
\title{APGKT: Exploiting Associative Path on Skills Graph for Knowledge Tracing 
\thanks{
Co-corresponding authors: Chenyang Bu (email: chenyangbu@hfut.edu.cn) and Fei Liu (email: feiliu@mail.hfut.edu.cn).\\
Chenyang Bu was supported in part by the National Natural Science Foundation of China under Grants 61806065 and 62120106008, and the Fundamental Research Funds for the Central Universities under Grant JZ2022HGTB0239. The source code is available at \url{https://github.com/DMiC-Lab-HFUT/APGKT-PRICAI2022}.}
}
\titlerunning{APGKT}
%
\author{Haotian Zhang
\inst{1}
\orcidID{0000-0003-0133-9762} \and
Chenyang Bu*
\inst{1}
\orcidID{0000-0001-8203-0956} \and \\
Fei Liu*
\inst{1,2}
\orcidID{0000-0003-0022-4103}
\and
Shuochen Liu
\inst{1}
\orcidID{0000-0003-4724-8989}
\and \\
Yuhong Zhang
\inst{1}
\orcidID{0000-0001-7031-0889}
\and
Xuegang Hu
\inst{1}
\orcidID{0000-0001-5421-6171}
}
\authorrunning{H. Zhang et al.}
%
\institute{Key Laboratory of Knowledge Engineering with Big Data (the Ministry of Education of China), School of Information Science and Computer Engineering, Hefei University of Technology, China \and
Jianzai Tech, Hefei, China }
%

%
\maketitle              
\begin{abstract}
    Knowledge tracing (KT) is a fundamental task in educational data mining that mainly focuses on students’ dynamic cognitive states of skills. The question-answering process of students can be regarded as a thinking process that considers the following two problems. One problem is which skills are needed to answer the question, and the other is how to use these skills in order. If a student wants to answer a question correctly, the student should not only master the set of skills involved in the question, but also think and obtain the associative path on the skills graph. The nodes in the associative path refer to the skills needed and the path shows the order of using them. 
    The associative path is referred to as the skill mode. Thus, obtaining the skill modes is the key to answering questions successfully. However, most existing KT models only focus on a set of skills, without considering the skill modes. We propose a KT model, called \mbox{APGKT}, that exploits skill modes. Specifically, we extract the subgraph topology of the skills involved in the question and combine the difficulty level of the skills to obtain the skill modes via encoding; then, through multi-layer recurrent neural networks, we obtain a student’s higher-order cognitive states of skills, which is used to predict the student’s future answering performance. Experiments on five benchmark datasets validate the effectiveness of the proposed model. 

\keywords{Educational data mining  \and knowledge tracing \and graph neural network.}
\end{abstract}
\section{Introduction}
\label{s1}
Recent advances in intelligent tutoring systems have promoted the development of online education and generated a large amount of online learning data \cite{liu2021survey,lf,liu2022tetci}. Knowledge tracing (KT) is used to model students' dynamic mastery of skills based on their historical learning data and to infer their future answering performance, which is a fundamental and essential task in computer-aided educational systems and online learning platforms \cite{tong2020hgkt,liu2021fuzzy}.

Bayesian knowledge tracing (BKT) \cite{BKT} was the first KT model proposed by Corbett et al. It models students' cognitive states using the hidden markov model (HMM) with limited representation capabilities \cite{pelanek2017bayesian}. Subsequently, deep learning models, such as deep knowledge tracing (DKT) \cite{DKT}, were developed, which model a student's learning process as a recurrent neural network (RNN), significantly improving the prediction performance of the traditional Bayesian-based KT. With the development of graph neural networks (GNN) \cite{EAAutoGL}, GNN-based KT models \cite{gkt,gikt}, which use the natural graph structure existing in skills to model students' cognition, have attracted considerable attention. Although KT models have developed rapidly in recent years, limitations still exist.

\begin{figure}[t]
    \centering
    \centerline{\includegraphics[scale=0.39]{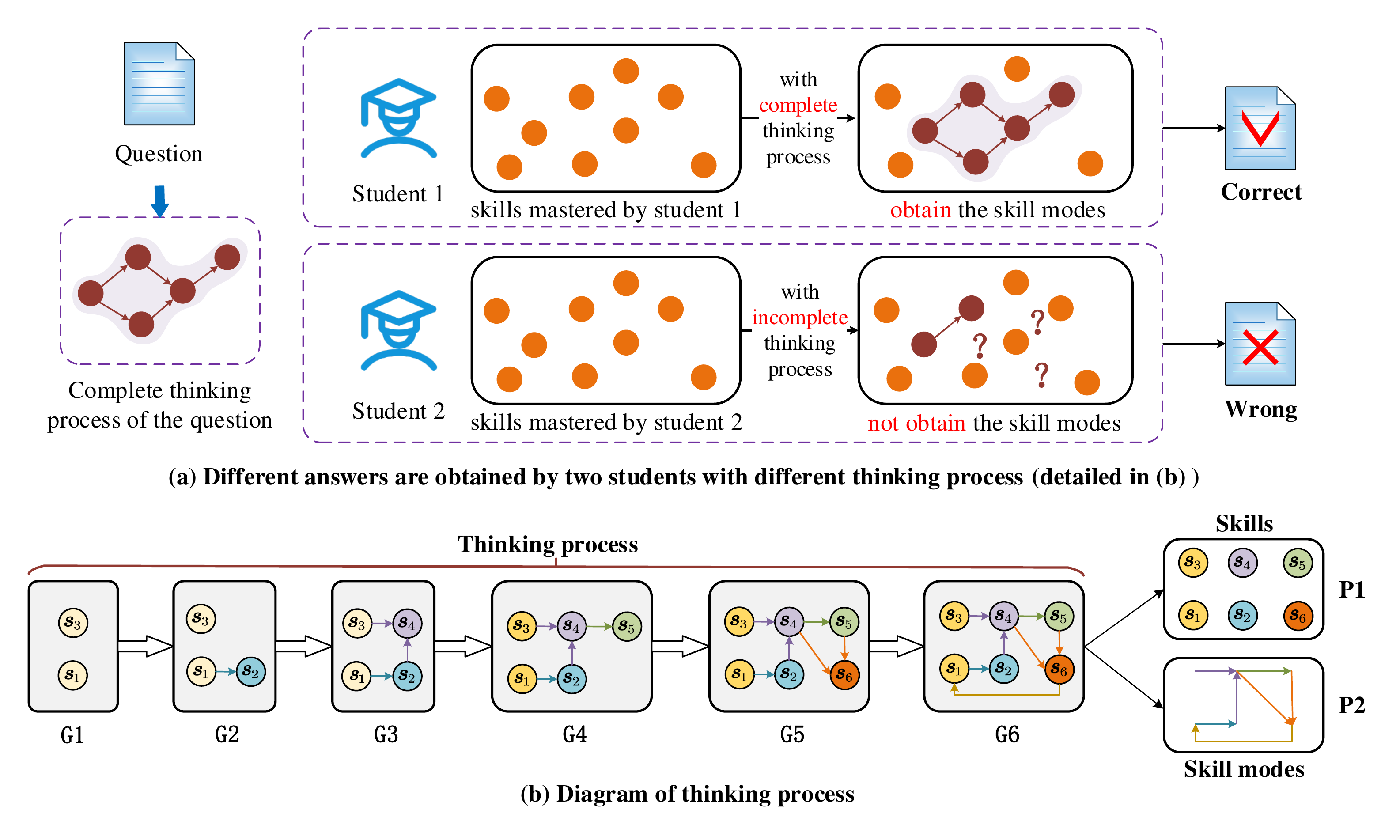}}
    \caption{(a) Instance of students answering questions. Given the same question, Student 1 and Student 2 provide different answers.
    Assuming that the skills mastery of the two students is similar, the student who cannot obtain the skill modes through thinking should have a higher probability of answering incorrectly. (b) Detailed thinking process of a student. $G_i$ represents every thinking state and the arrow connecting two states represents a state transition, indicating a student's thinking and associative behavior.}
    \label{fig:demo}
\end{figure}

Most of the existing KT models assume that students could obtain the correct answer only if they mastered all the skills; therefore, they use the cognitive state of the skills to predict a student’s future answering performance. However, they ignore the thinking process of students. In addition to mastering skills, two points need to be considered to predict the future answering performance of a student: (1) finding the skills needed to answer a question among all the skills mastered, and (2) obtaining a reasonable order of use for these skills. If a student wants to answer a question correctly, the student should not only master the set of skills involved in the question but should also think and obtain the associative path on the skills graph, the nodes in which are the skills to be used, and the path showing the order of using them. Here, the associative path is referred to as the skill mode. 
If students only master the skills (e.g., $P1$ in Fig. \ref{fig:demo}(b)), the students cannot solve the problem because they may not establish an association between $s_1$ and $s_2$; they do not think of using $s_2$ to solve the problem. At this time, the students get stuck in processing the association from $G1$ to $G2$ shown in Fig. \ref{fig:demo}(b). Students may fail to establish an association between $s_2$, $s_3$, and $s_4$ as well. At this time, the student gets stuck in processing the association from $G2$ to $G3$ shown in Fig. \ref{fig:demo}(b). Students who do not master any of the processes in $P2$ may fail to solve the problem. 
Thus, obtaining skill modes is the key to answering questions successfully. As shown in Fig. \ref{fig:demo}(a), Student 1 and Student 2 provide different answers for the same question. Assuming that the skill mastery of the two students is similar, the student who cannot obtain the skill modes through thinking should have a higher probability of answering incorrectly (as shown in Fig. \ref{fig:demo}(a)). Students must use the skills they have mastered, the information in the question, and their experience to find the skills needed to answer a question and convert the thinking process into answers (as shown in Fig. \ref{fig:demo}(b)). This study assumed that students will have a higher probability of getting a question wrong if they only master the skills without mastering the skill modes.

\mbox{APGKT} is proposed considering skill modes (e.g., $P2$ in Fig. \ref{fig:demo}(b)) to improve performance of KT.
The main contributions of this study are as follows:
\begin{itemize}
    \item
    This study exploits the associative path on the skills graph for knowledge tracing (KT). The thinking process (i.e., obtaining the associative path) has been demonstrated to be indispensable for achieving a correct answer (detailed in Fig. \ref{fig:demo}). However, most of the existing KT models only consider whether the set of skills involved in the question have been mastered when predicting a student’s future answering performance.
    \item
    The proposed APGKT model includes the concept of skill modes and higher-order cognitive states. Considering the dynamic process of students thinking and answering questions, the skills associated with a specific problem are considered as a whole to consider the organizational association. We combine the cognitive state of the skills and the skill modes into a higher-order cognitive state to accurately represent the cognitive processes of students.
    \item
    Extensive experiments on five public datasets proved that the prediction results of our model are better than those of baseline models, owing to the consideration of the thinking process during KT.
\end{itemize}


\section{Related Work}
\label{s2}
In this section, related work regarding KT and the existing GNN-based KT models is introduced. 
\subsection{Knowledge Tracing}\label{s2.1}
\mbox{KT} as a student modeling technique has attracted extensive research work. 
Existing \mbox{KT} models can be divided into three main categories: probabilistic models, logistic models, and deep learning-based models \cite{liu2021survey}.
(1) \textbf{Probabilistic models}, which assume a Markov process to represent the learning process of students, are mainly of two types \cite{liu2021survey}: \mbox{BKT} \cite{BKT} and \mbox{DBKT} \cite{DBKT}. They use unobservable nodes in the HMM to represent the knowledge state, and Bayesian networks and dynamic Bayesian networks for KT.
(2) \textbf{Logistic models}, which assume that the probability of correctly answering questions can be expressed as a mathematical framework of students and skills parameters, are mainly of three types \cite{liu2021survey}: \mbox{LFA} \cite{LFA}, \mbox{PFA} \cite{PFA}, and \mbox{KTM} \cite{KTM}. They use the output of the logistic regression function to represent the knowledge state, and logistic regression or factorization machines to model the knowledge state change.
(3) \textbf{Deep learning-based models} adapt to complex learning processes, especially in the face of extensive interactive data \cite{liu2021survey}, are being considered. Deep learning is a powerful tool to implement nonlinearity and feature extraction. \mbox{DKT} \cite{DKT}, the first deep learning-based model for KT, uses a RNN to model the cognitive state of students and has achieved excellent results. Subsequently, this model has been further developed into memory-aware \cite{DKVMN}, problem-aware \cite{SimEx,QuesNet,liu2019ekt}, and attention \cite{SAKT,FAKTWANG2021223,InterpretableAKT,SAINT+} models \cite{liu2021survey}, which use the interactive information in students' responses. Due to the natural graph structure of the KT task, GNN-based KT models have attracted researchers (detailed in Section \ref{s2.2}).

\subsection{GNN-based KT models}\label{s2.2}
GNNs, which process complex graph-structured data, have developed rapidly in recent years. In GNNs, a graph is a data structure that models a set of objects (nodes) and their relationships (edges). From the perspective of data structure, graph structures naturally exist within skills \cite{gkt}. Therefore, combining the graph structure of the components (such as skills or questions) with relational inductive bias should improve the performance of KT models \cite{liu2021survey}.

Recently, several \mbox{KT}-structure frameworks based on GNNs have been developed. For example, GKT \cite{gkt} conceptualizes the underlying graph structure of skills into a graph to influence the updating process of the cognitive states of skills. HGKT \cite{tong2020hgkt} mines the hidden hierarchical relationships among exercises by constructing a hierarchical exercise graph. GIKT \cite{gikt} aggregates the embedding of questions and skills through a graph convolutional network (GCN) to extract the higher-order information from them. By introducing the transfer of knowledge \cite{transflr}, SKT \cite{tong2020structure} further explores the knowledge structure and captures multiple relations in it to model the influence propagation among concepts. JKT \cite{song2021jkt} captures high-level semantic information and improves model interpretability by modeling the multi-dimensional relationships of ``exercise-to-exercise'' and ``concept-to-concept'' as graphs and fusing them with the ``exercise-to-concept'' relationship. Most existing GNN-based KT models only consider the graph structure within the set of skills involved in questions (e.g. $P1$ in Fig. \ref{fig:demo}(b)). Therefore, they lack the mining and utilization of information in the skill modes (e.g., $P2$ in Fig. \ref{fig:demo}(b)), which is what we focused on in this study. 

\subsection{GIKT}\label{s3}
Our work is inspired by a graph-based interaction model for knowledge tracing (GIKT), and we refer readers to the reference \cite{gikt} for more details about GIKT.

\begin{figure}[t]
    \centering
    \includegraphics[scale=0.5]{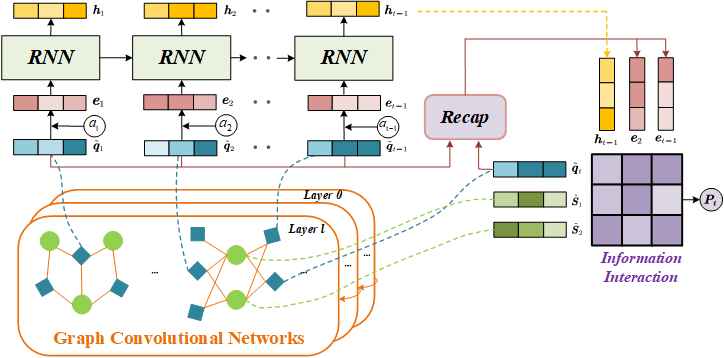}
    \caption{Framework of the GIKT \cite{gikt} model.}
    \label{fig:gikt}
\end{figure}

\subsubsection{Embedding Propagation}\label{s3.3}
GIKT models the relationship between questions and skills as a bigraph and uses multiple layers of GCN to aggregate their embeddings. After the GCN embedding propagation and aggregation processes, higher-order questions and skill-embedding representations $\tilde{q}$ and $\tilde{s}$ are obtained, respectively.

\subsubsection{Student State Evolution}\label{s3.4}

For each historical time $t$, GIKT obtains a representation of exercise $e_t$ by concatting the embeddings of aggregated question $\tilde{q_t}$ and answer $a_t$. Then a long short-term memory network (LSTM) is used to learn the changes in the cognitive states $h_t$ of students using $e_t$ as input.

\subsubsection{History Recap Module}\label{s3.5}

GIKT uses a history recap module to select the history exercises related to the current answered questions to better represent the student’s ability to answer the current specific question $q_t$. GIKT provides two methods for selecting history exercises $I_e$: hard and soft selections. The hard selection method only selects questions with skills identical to the current answered question each time and the soft selection method uses the similarity between the questions to select the top $k$-related problems with the highest correlations with the current question being answered.

\subsubsection{Generalized Interaction Module}\label{s3.6}

In this module, GIKT uses $\left \langle h_t,\tilde{q_t} \right \rangle$ to indicate the student's mastery of question $q_t$, $\left \langle h_t,\tilde{s_j} \right \rangle$ to indicate the student's mastery of related skill $s_j \in \mathcal{N}_{q_t}$, $\left \langle h_i,\tilde{q_t} \right \rangle$, and $\left \langle h_i,\tilde{s_j} \right \rangle$ to represent the interaction of the current student state with historical states. GITK considers the interaction information of all these states to obtain the predicted value.

\section{APGKT: Proposed Model}\label{s4}
In this section, we introduce the framework (detailed in Section \ref{s4.2}) of our model, which includes graph construction and representation (detailed in Section \ref{s4.3}), and student state evolution and prediction (detailed in Section \ref{s4.4}).

\begin{figure}[t]
    \centering
    \centerline{\includegraphics[scale=0.325]{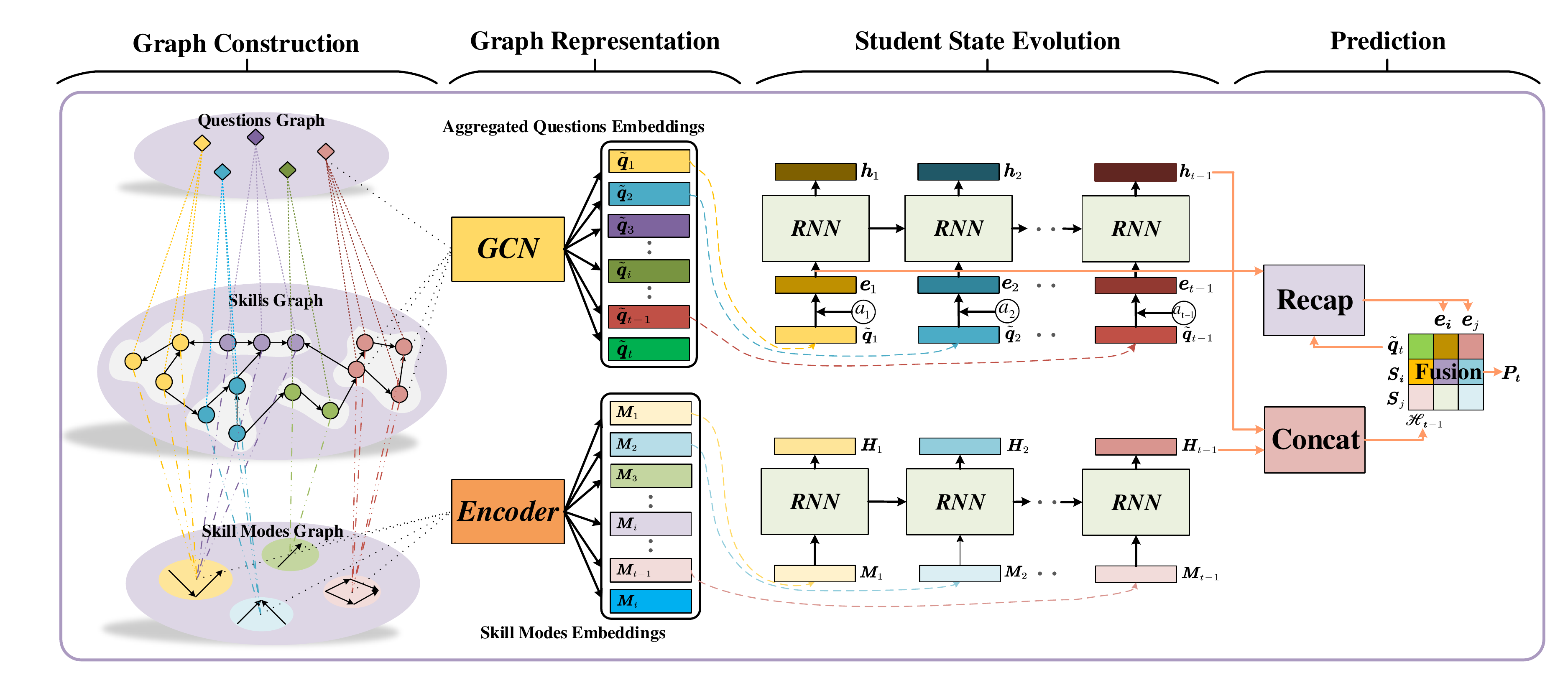}}
    \caption{Complete framework of the \mbox{APGKT} model. The first module on the left is the graph construction, the bottom of which is the skill modes graph we are concerned about. The next module is graph representation, where the efficient representation of the questions and the skill modes are obtained. In the student state evolution module, we obtain a student’s cognitive state of skills and skill modes. Finally, the prediction module obtains the final prediction by fusing the higher-order cognitive state obtained by $Concat$ and other state information.}
    \label{fig:APGKT}
\end{figure}

\subsection{Framework}\label{s4.2}

The framework of the \mbox{APGKT} model is shown in Fig. \ref{fig:APGKT}. First, we construct a graph and obtain its representations (detailed in Section \ref{s4.3}). We then obtain a student’s higher-order cognitive states by splicing the cognitive state of skills and skill modes, which is then used to predict the performance of the student (detailed in Section \ref{s4.4}). In the following sections, we describe in detail each module of our model.

\subsection{Graph Construction and Representation}\label{s4.3}
The structure of the graph is first described. Then, the construction of skill graph, the generation and representation of skill modes are detailed.

\subsubsection{Structure of the Graph}\label{s4.3.1}

To represent the relationship between questions, skills, and skill modes, we constructed a graph with three layers for three relationships (as shown in Fig. \ref{fig:APGKT}).

\textbf{(1) Three layers.}
\textbf{a)} The top layer is a question graph that contains all the questions from the student’s answer record. We represent these questions by $Q = \left\{q_1,q_2,...,q_{n_q}\right\}$, where $n_q$ denotes the total number of questions. 
\textbf{b)} The middle layer is a skills graph, which consists of the skills involved in all the questions. These skills are represented as $S = \left\{s_1,s_2,...,s_{n_s}\right\}$, where $n_s$ is the total number of skills. 
\textbf{c)} The bottom layer is a skill-mode graph, which contains all the obtained skill modes.

\textbf{(2) Three relations.}
\textbf{a)} Each question $q_i$ in the question graph is associated with a skill set in the skills graph, and we represent this skill set as ${Sset}_i = \left\{s_1^i,s_2^i,...,s_{h_i}^i\right\}, h_i \ge 1$. $s_1^i,s_2^j,...,s_{h_i}^i$ are skills related to question $q_i$ and $h_i$ indicates the number of skills related to question $q_i$. The skills in the skills graph are divided into several skill sets based on the questions. The relationship between questions $Q$ and skills $S$ is represented by a matrix QS. QS is a two-dimensional matrix of size $n_q \times n_s$, where $QS_{i,j} = 1$ indicates that $q_i$ is related to $s_j$.
\textbf{b)} The relationship between skills is constructed using several methods, which will be introduced in Section \ref{s4.3.2}. This relationship is represented by a two-dimensional adjacency matrix SS of size $n_s \times n_s$, where $n_s$ is the number of skills. $0 \leq SS_{i,j} \leq 1$ indicates the strength of the association between $s_i$ and $s_j$. Note that $SS_{i,j}$ and $SS_{j,i}$ represent different relationships between the skills. 
\textbf{c)} The method for obtaining the relationship between skills and skill modes is introduced in Section \ref{s4.3.3}. A skill may belong to different skill sets and different skill modes because it is simultaneously associated with different questions, and the number of skill modes equals the number of skill sets, as shown in Fig. \ref{fig:APGKT}.

\subsubsection{Skills Graph Construction}\label{s4.3.2}

APGKT needs to use the graph structure of skills when evaluating a student’s proficiency in skills and skill modes. However, in most cases, the structure of the skills is not explicitly provided. Nakagawa et al. \cite{gkt} introduced statistics-based and learning-based approaches for implementing the latent graph structure, of which the former are more efficient with less time consumption (detailed in Table 2 in \cite{gkt}). From the aspect of statistics-based approaches, we assumed that the higher the frequency of two skills appearing together in the same question, the stronger the strength of the association between the two skills. This was not considered in the statistics-based approaches in \cite{gkt}. Therefore, a frequency-based method is proposed in this subsection.

\textbf{Frequency-based method} generates a connected graph according to the number of times two skills appear together in the same question and the number of times two skills appear separately in different questions. This is calculated using Eq. \eqref{freq}.
\begin{align}
    \label{freq} SS_{i,j} = \frac{n_{i,j}}{\sum_{k=1}^{n_s}n_{i,k}},
\end{align}
where $n_{i,j}$ represents the times two skills appear together in the same question.


\subsubsection{Skill Modes Generation and Representation}\label{s4.3.3}

Through the complete thinking process, the skill modes are obtained, which represent the associative paths on the skills graph (as shown in Fig. \ref{fig:demo}(b)). In this subsection, the generation and representation of the skill modes are designed.

Considering that students usually have a thinking process from easy to difficult when answering questions, we obtain an effective representation of the skill modes using the encoded association paths and difficulty levels of skills. Specifically, we first obtain the difficulty level of all the skills through statistical information using Eq. \eqref{diff}.
Then, we obtain the ascending subscripts of the skills in $Sset_i$ according to the skill difficulty and referred to $Idx_i = \left\{i,j,...,k \right\}$. We finally extract the local topological structure of $Sset_i$ in the SS using Eq. \eqref{m_i}. That is, the values of the ${i,j,...,k}$th row and  ${i,j,...,k}$th column in the SS are extracted and flattened to obtain the initial representation $m_i$ of the skill mode. 
\begin{align}
    \label{diff} Diff_{s_i} &= \frac{n^i}{N^i},
\end{align}
where $n^i$ is the number of wrong answers to questions containing skill $s_i$ and $N^i$ is the number of questions containing skill $s_i$.
\begin{align}
    \label{m_i} m_i &= Flatten(\sum_{i' \in Idx_i} \sum_{j' \in Idx_i} SS_{i',j'}),
\end{align}
where Flatten indicates making multidimensional data one-dimensional.


We encode the initial representation of the skill modes through an encoder module to obtain the embedding of the skill modes $M_i$ (Eq. \eqref{M_i}), and then calculate the mean squared error (mse) with the encoded $m_i$ after decoding it to obtain the reconstruction loss $Reloss$ using Eq. \eqref{reloss}. Finally, we minimize $Reloss$ to obtain an effective representation of the skill modes.
\begin{align}
    \label{M_i}&  M_i = \sigma (W_M \times m_i + b_M) \\
    \label{reloss}& Reloss = \frac{1}{n_q} \sum_1^{n_q}(M_i - m_i)^2 
\end{align}

In Eq. \eqref{M_i}, $\sigma$ indicates a nonlinear mapping, and $W_M$ and $b_M$ indicate the weights and biases, respectively, in the encoder that will be trained.

\subsection{Student State Evolution and Prediction}\label{s4.4}

For each time step $t$, the embedding of the aggregated question $\tilde{q_t}$ and skill modes $M_t$ of $q_t$ are provided as inputs into the LSTM to learn a student’s mastery of skills and skill modes. Next, we connect the cognitive states of the student’s skills and skill modes through the $Concat$ module to obtain the student’s higher-order cognitive state ${\mathcal{H}_t}$ using Eq. \eqref{gaoh}. Finally, we incorporate the student’s higher-order cognitive state ${\mathcal{H}_t}$ in \eqref{gaoh} to improve the prediction of GIKT, and obtain the final prediction $p_t$ as shown in Eq. \eqref{alpha} and Eq. \eqref{p_t} \cite{gikt}.
\begin{align}
    \label{gaoh} {\mathcal{H}_t} &= [h_t, H_t],
\end{align}
where $[\cdot]$ represents vector concatenation.
\begin{align}
    \label{alpha} \alpha_{i,j} &= Softmax_{i,j}(W^T[f_i,f_j] + b), \\
    \label{p_t} p_t &= \sum_{f_i \in I_e \cup \left\{\mathcal{H}_t \right\}} \sum_{f_j \in \tilde{\mathcal{N}_{q_t}} \cup \left\{\tilde{q_t} \right\}} \alpha_{i,j} g(f_i,f_j),
\end{align}
where $p_t$ indicates the predicted result at time t, $I_e$ indicates history exercises related to the $q_t$. ${\mathcal{H}_t}$ is the higher-order cognitive state of the student. $\tilde{\mathcal{N}_{q_t}}$ is the aggregated neighbor skill embedding of $q_t$. $g$ represents the inner product.

APGKT is optimized by minimizing the cross-entropy loss between the predicted and the true values using gradient descent as shown in Eq. \eqref{loss}.
\begin{equation}\label{loss}
    \mathcal{L} = - \sum_t(a_t \log p_t + (1-a_t)\log(1-p_t)),
\end{equation}
where $a_t$ represents the true value of the students' answer at time t.

\section{Experiments}\label{s5}
Experiments are conducted on five real-world datasets to demonstrate the effectiveness of the proposed model. First, the setup is introduced, including the datasets, baselines, and implementation details. Then, the comparing results and Nemenyi tests are presented. Finally, the parameters in the model are analyzed.

\subsection{Setup}
The setup of the experiments is introduced, including the five datasets, the compared baselines, and the implementation details.

\subsubsection{Datasets}\label{s5.1}
Five real-world datasets were used and their statistics are listed in Table \ref{datasets}.
To verify the effectiveness of our model in the multi-skills scenario, we further processed the assist09 dataset, and only retained the questions involving multiple skills and students' answer records to form the dataset assist09-muti. 
The questions in CSEDM, FrcSub, Math1, and Math2 were all related to more than one skill, and there were no questions related to a single skill.

\begin{table*}[t]\label{t1}
\centering
\caption{Dataset statistics} \label{datasets}
\begin{tabular}{lllllll}
\hline
Datasets            & assist09 & assist09-muti & CSEDM & FrcSub & Math1 & Math2 \\ \hline
Number of students  & 3002     & 1793          & 343   & 536    & 4209  & 3911  \\
Number of questions & 17705    & 3014          & 236   & 20     & 15    & 16    \\
Number of skills    & 123      & 54            & 18    & 8      & 11    & 16    \\ \hline
\end{tabular}
\end{table*}

\subsubsection{Comparison Baselines}\label{s5.2}
To verify the effectiveness of our model, APGKT is compared with the following baselines: DKT \cite{DKT}, DKVMN \cite{DKVMN}, GKT \cite{gkt}, GIKT \cite{gikt} (detailed in Section \ref{s3}).


\begin{table*}[t]
\centering
\caption{Comparison in terms of AUC} \label{auc}
\begin{tabular}{ccccccc}
\hline
Dataset & DKT\cite{DKT}  & DKVMN\cite{DKVMN} & GKT\cite{gkt}  & GIKT\cite{gikt}   & APGKT (Our model)        \\ \hline
assist09      & 0.6995 & 0.7112 & 0.7230 & 0.7742 & \textbf{0.7767} \\
assist09-muti & 0.6961 & 0.7106 & 0.7320 & 0.7763 & \textbf{0.7817} \\
CSEDM         & 0.7543 & 0.7626 & 0.7647 & 0.7836 & \textbf{0.7902} \\
FrcSub        & 0.8891 & 0.8729 & 0.8748 & 0.8982 & \textbf{0.9059} \\
Math1         & 0.8349 & 0.8403 & 0.8456 & 0.8892 & \textbf{0.8922} \\
Math2         & 0.8084 & 0.8159 & 0.8181 & 0.8681 & \textbf{0.8695} \\ \hline
\end{tabular}
\end{table*}

\subsubsection{Implementation Details}\label{s5.3}
The APGKT code was written using TensorFlow. The datasets were divided into training and testing sets in the ratio of 8:2. We set the length of skills, questions, and answer embeddings to 100, which were not pretrained but were randomly initialized and then optimized during training. The relationship between the skills was constructed using the Frequency-based method (detailed in Section \ref{s4.3.2}). Finally, we used the Adam optimizer with a learning rate of 0.003 to optimize all the trainable parameters.

\subsection{Results}\label{s5.4}
Results including the mean AUC results, the Nemenyi tests, and the parametric analysis are illustrated in this subsection.

\subsubsection{Comparison in terms of AUC}
We used AUC as the evaluation criterion, and Table \ref{auc} shows the AUC scores of the baseline models and our model. We observed that the AUC scores of APGKT were the highest (denoted in bold) for all the datasets, which demonstrates the effectiveness of the proposed method.  On comparing the AUC scores of the models on the assist09 and assist09-muti datasets, we observed that our model performed better than the baseline models in multi-skill scenarios. This may be due to the abundant skill modes available in our model in multi-skill scenarios, which improves its predictive performance.

\subsubsection{Nemenyi Test}

\begin{figure}[t]
    \centering
    \centerline{\includegraphics[scale=0.58]{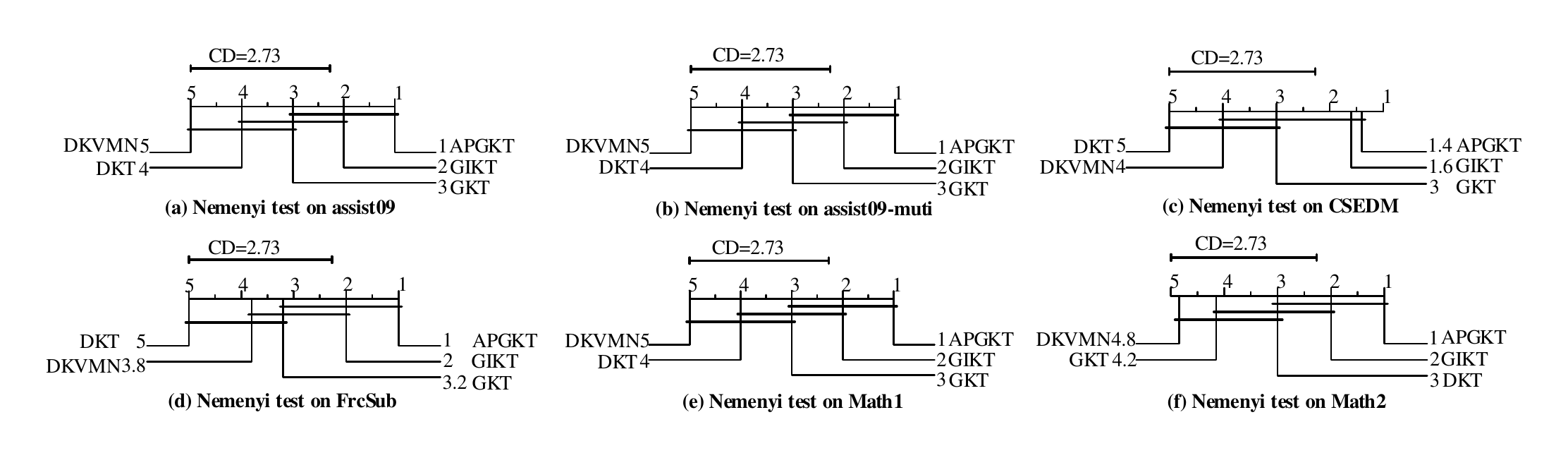}}
    \caption{Nemenyi test results of the proposed model and baselines. The results demonstrate the better performance of the proposed model.}
    \label{fig:Nemenyi}
\end{figure}

In the experiments, Nemenyi tests \cite{Nemenyi} were conducted to statistically compare the five algorithms over five datasets (as shown in Fig. \ref{fig:Nemenyi}). The test results showed that our model performed better than other models. 

\subsubsection{Parametric Analysis}

\begin{figure}[t]
    \centering
    \centerline{\includegraphics[scale=0.6]{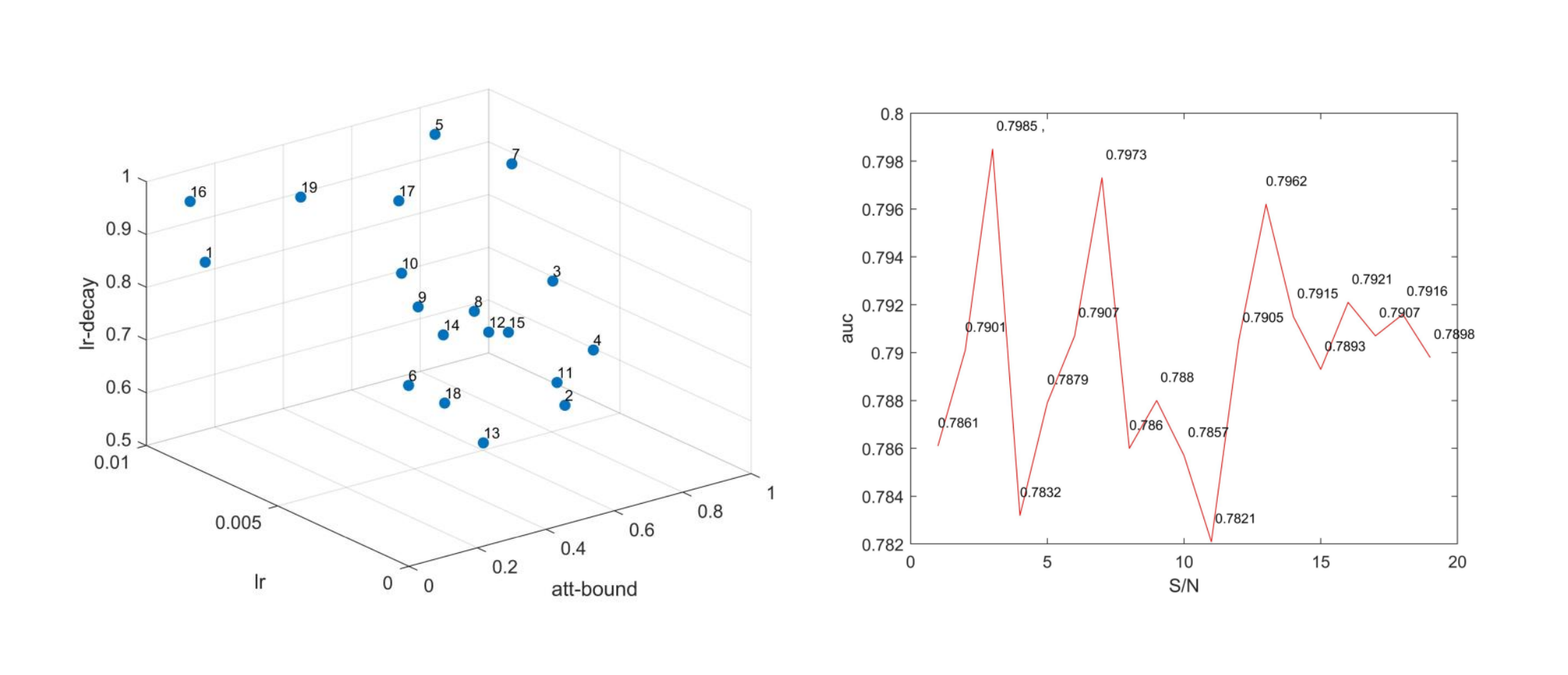}}
    \caption{Parameter analysis for APGKT. It is observed that our model outperforms the baselines although the parameters underwent constant changes.}
    \label{fig:Parametric}
\end{figure}

We also conducted parameter analyses on the CSEDM dataset to analyze the model's sensitivity to the parameters. Bayes opt (\url{https://github.com/fmfn/BayesianOptimization}) was used to tune the learning rate of Adam optimizer (lr), learning rate decay (lr-decay), and threshold for determining whether two questions are related (att-bound). They were initialized ranging from 0 to 1. It is observed that the performance of our model was superior to that of the baseline models although the parameters underwent constant changes.

\section{Conclusion}\label{s6}

Most of the existing KT models ignore the thinking process between specific skills, leading to suboptimal prediction performance. We introduced skill modes and higher-order cognitive states to solve this problem and proposed a novel model named APGKT. Specifically, we considered the dynamic process of students thinking and answering questions, and further explored the relationship between the specific skills involved in the questions. Extensive experiments on five public datasets verified that the proposed model outperformed the baseline models. Since the thinking process of students is actually a complex cognitive process, which is affected by many factors such as psychology, in the future, we will further explore the representation and application of the thinking process to improve the model.

\bibliographystyle{elsarticle-num}
\bibliography{bibliography}        

\begin{thebibliography}{10}
\expandafter\ifx\csname url\endcsname\relax
  \def\url#1{\texttt{#1}}\fi
\expandafter\ifx\csname urlprefix\endcsname\relax\def\urlprefix{URL }\fi
\expandafter\ifx\csname href\endcsname\relax
  \def\href#1#2{#2} \def\path#1{#1}\fi

\bibitem{liu2021survey}
Q.~Liu, S.~Shen, Z.~Huang, E.~Chen, Y.~Zheng, A survey of knowledge tracing,
  arXiv preprint arXiv:2105.15106 (2021).

\bibitem{lf}
X.~Hu, F.~Liu, C.~Bu, Research advances on knowledge tracing models in
  educational big data, Journal of Computer Research and Development 57~(12)
  (2020) 2523--2546.

\bibitem{liu2022tetci}
C.~Bu, F.~Liu, Z.~Cao, L.~Li, Y.~Zhang, X.~Hu, W.~Luo, Cognitive diagnostic
  model made more practical by genetic algorithm, IEEE Transactions on Emerging
  Topics in Computational Intelligence (TETCI) (2022).

\bibitem{tong2020hgkt}
H.~Tong, Z.~Wang, Q.~Liu, Y.~Zhou, W.~Han, {HGKT}: Introducing hierarchical
  exercise graph for knowledge tracing, arXiv preprint arXiv:2006.16915 (2020).

\bibitem{liu2021fuzzy}
F.~Liu, X.~Hu, C.~Bu, K.~Yu, Fuzzy {B}ayesian knowledge tracing, IEEE
  Transactions on Fuzzy Systems (TFS) 30~(7) (2022) 2412--2425.

\bibitem{BKT}
A.~T. Corbett, J.~R. Anderson, Knowledge tracing: Modeling the acquisition of
  procedural knowledge, User Modeling and User-adapted Interaction 4~(4) (1994)
  253--278.

\bibitem{pelanek2017bayesian}
R.~Pel{\'a}nek, Bayesian knowledge tracing, logistic models, and beyond: An
  overview of learner modeling techniques, User Modeling and User-Adapted
  Interaction 27~(3) (2017) 313--350.

\bibitem{DKT}
C.~Piech, J.~Spencer, J.~Huang, S.~Ganguli, M.~Sahami, L.~Guibas,
  J.~Sohl-Dickstein, Deep knowledge tracing, in: Proceedings of International
  Conference on Neural Information Processing Systems (NeurIPS), 2015, pp.
  505--513.

\bibitem{EAAutoGL}
C.~Bu, Y.~Lu, F.~Liu, Automatic graph learning with evolutionary algorithms: An
  experimental study, in: Proceedings of The Pacific Rim International
  Conference on Artificial Intelligence 2021 (PRICAI): Trends in Artificial
  Intelligence, Springer International Publishing, Cham, 2021, pp. 513--526.

\bibitem{gkt}
H.~Nakagawa, Y.~Iwasawa, Y.~Matsuo, Graph-based knowledge tracing: Modeling
  student proficiency using graph neural network, in: Proceedings of
  IEEE/WIC/ACM International Conference on Web Intelligence (WI), IEEE, 2019,
  pp. 156--163.

\bibitem{gikt}
Y.~Yang, J.~Shen, Y.~Qu, Y.~Liu, K.~Wang, Y.~Zhu, W.~Zhang, Y.~Yu, {GIKT}: A
  graph-based interaction model for knowledge tracing, in: Proceedings of
  Machine Learning and Knowledge Discovery in Databases, Springer International
  Publishing, Cham, 2021, pp. 299--315.

\bibitem{DBKT}
T.~Käser, S.~Klingler, A.~G. Schwing, M.~Gross, Dynamic bayesian networks for
  student modeling, IEEE Transactions on Learning Technologies 10~(4) (2017)
  450--462.

\bibitem{LFA}
H.~Cen, K.~Koedinger, B.~Junker, Learning factors analysis-a general method for
  cognitive model evaluation and improvement, in: Proceedings of Intelligent
  Tutoring Systems, Springer Berlin Heidelberg, Berlin, Heidelberg, 2006, pp.
  164--175.

\bibitem{PFA}
P.~I. Pavlik, H.~Cen, K.~R. Koedinger, Performance factors analysis-a new
  alternative to knowledge tracing, in: Proceedings of Conference on Artificial
  Intelligence in Education: Building Learning Systems That Care: From
  Knowledge Representation to Affective Modelling, IOS Press, NLD, 2009, pp.
  531--538.

\bibitem{KTM}
J.-J. Vie, H.~Kashima, Knowledge tracing machines: Factorization machines for
  knowledge tracing, in: Proceedings of the AAAI Conference on Artificial
  Intelligence (AAAI), Vol.~33, 2019, pp. 750--757.

\bibitem{DKVMN}
J.~Zhang, X.~Shi, I.~King, D.~Y. Yeung, Dynamic key-value memory networks for
  knowledge tracing, in: Proceedings of International Conference on World Wide
  Web (WWW), ACM, 2017, pp. 765--774.

\bibitem{SimEx}
Q.~Liu, Z.~Huang, Z.~Huang, C.~Liu, E.~Chen, Y.~Su, G.~Hu, Finding fimilar
  exercises in online education systems, in: Proceedings of ACM SIGKDD
  International Conference on Knowledge Discovery \&amp; Data Mining, KDD '18,
  Association for Computing Machinery, New York, NY, USA, 2018, pp. 1821--1830.

\bibitem{QuesNet}
Y.~Yin, Q.~Liu, Z.~Huang, E.~Chen, W.~Tong, S.~Wang, Y.~Su, {QuesNet}: A
  unified representation for heterogeneous test questions, in: Proceedings of
  ACM SIGKDD International Conference on Knowledge Discovery \&amp; Data
  Mining, KDD '19, Association for Computing Machinery, New York, NY, USA,
  2019, pp. 1328--1336.

\bibitem{liu2019ekt}
Q.~Liu, Z.~Huang, Y.~Yin, E.~Chen, H.~Xiong, Y.~Su, G.~Hu, {EKT}:
  Exercise-aware knowledge tracing for student performance prediction, IEEE
  Transactions on Knowledge and Data Engineering (TKDE) 33~(1) (2019) 100--115.

\bibitem{SAKT}
S.~Pandey, G.~Karypis, A self-attentive model for knowledge tracing, CoRR
  abs/1907.06837 (2019).
\newblock \href {http://arxiv.org/abs/1907.06837} {\path{arXiv:1907.06837}}.

\bibitem{FAKTWANG2021223}
X.~Wang, X.~Mei, Q.~Huang, Z.~Han, C.~Huang, Fine-grained learning performance
  prediction via adaptive sparse self-attention networks, Information Sciences
  545 (2021) 223--240.

\bibitem{InterpretableAKT}
J.~Zhu, W.~Yu, Z.~Zheng, C.~Huang, Y.~Tang, G.~P.~C. Fung, Learning from
  interpretable analysis: Attention-based knowledge tracing, in: Proceedings of
  Artificial Intelligence in Education, Springer International Publishing,
  Cham, 2020, pp. 364--368.

\bibitem{SAINT+}
D.~Shin, Y.~Shim, H.~Yu, S.~Lee, B.~Kim, Y.~Choi, {SAINT+}: {I}ntegrating
  temporal features for ednet correctness prediction, in: Proceedings of LAK21:
  International Learning Analytics and Knowledge Conference, LAK21, Association
  for Computing Machinery, New York, NY, USA, 2021, pp. 490--496.

\bibitem{transflr}
J.~M. Royer, Theories of the transfer of learning, Educational Psychologist
  14~(1) (1979) 53--69.

\bibitem{tong2020structure}
S.~Tong, Q.~Liu, W.~Huang, Z.~Huang, E.~Chen, C.~Liu, H.~Ma, S.~Wang,
  Structure-based knowledge tracing: An influence propagation view, in:
  Proceedings of IEEE International Conference on Data Mining (ICDM), IEEE,
  2020, pp. 541--550.

\bibitem{song2021jkt}
X.~Song, J.~Li, Y.~Tang, T.~Zhao, Y.~Chen, Z.~Guan, {JKT}: A joint graph
  convolutional network based deep knowledge tracing, Information Sciences 580
  (2021) 510--523.

\bibitem{Nemenyi}
J.~Dem{\v{s}}ar, Statistical comparisons of classifiers over multiple data
  sets, The Journal of Machine Learning Research 7 (2006) 1--30.

\end{thebibliography}

%

\end{document}